\documentclass{ws-ijqi}

\begin{document}

\markboth{F. Gentile, A. Montorsi and M. Roncaglia}{Entanglement generation and dynamics for a Bose-Hubbard model in a double-well potential.}

%
\catchline{}{}{}{}{}
%

\title{ENTANGLEMENT GENERATION AND DYNAMICS FOR A BOSE-HUBBARD MODEL 
IN A DOUBLE-WELL POTENTIAL}

\author{F. Gentile}

\address{INRIM, strada delle Cacce 91, 10135 Torino, Italy\\
fabio.gentile.88@gmail.com}

\author{A. Montorsi}

\address{Dipartimento di Fisica del Politecnico, corso Duca degli Abruzzi
24, 10129 Torino, Italy\\
arianna.montorsi@polito.it}

\author{M. Roncaglia}

\address{INRIM, strada delle Cacce 91, 10135 Torino, Italy\\
marco.roncaglia.it@gmail.com}

\maketitle

\begin{history}
\received{(Day Month Year)}
\revised{(Day Month Year)}
\end{history}

\begin{abstract}
The study of entanglement between bosonic systems is of primary importance for establishing feasible resources 
needed for implementing quantum information protocols, both in their interacting atomic or photonic realizations. 
Atomic systems are particularly efficient in the production of large amounts of entanglement, providing 
higher information density than conventional qubit entangled states. Such increased quantum resources pave the way to 
novel fundamental tests of nature and efficient applications in quantum information, metrology and sensing.     
We consider a basic setup made up of two parties A and B, each one populated by a single level bosonic variable. 
The bosons are interacting and can hop between A and B, thus describing a two-site Bose-Hubbard Hamiltonian.  
We consider the generation of quantum states in several situations that cover the majority of physical realizations: 
ground state, finite temperature, unitary dynamics, dissipation through dephasing and loss of particles. 
The system is analyzed through truncated exact diagonalization, as a function of the microscopic parameters.
The non separability of the obtained quantum states is estimated by means of the negativity, 
which has recently been proven to be a suitable measure of entanglement\cite{RoMoGe2013}. 
Finally, we calculate lower bounds of the entanglement of formation, an indicator that quantifies the minimal amount of 
entanglement resources needed to build up such states.
\end{abstract}


\section{Introduction} 	

In the last decades the concept of quantum entanglement acquired even more importance in various 
branches of modern science\cite{Genovese_review05,AmFaOsVe2008}.  
One of its most important applications stands in quantum information, where its usage is fundamental in several protocols. 
The entanglement of quantum states used in quantum information protocols is consumed by them, 
usually by measure operations. Therefore, entanglement represents a resource\cite{BlDaZw2008,NiCh2000} and its measure becomes crucial.   
Unfortunately, the definition of an entanglement measure is not a trivial task for general mixed states. 
Nevertheless, there are some special cases in which a suitable entanglement measure can be defined\cite{BeZy2006}. 
In this article we study a simple system for which a suitable entanglement measure has very recently been proven to exist: 
a fixed number of bosons that occupy two spatially separated quantum states and have a finite probability to hop between them. 
This system is readily implementable in the lab in two different ways, at least. 
The first is using ultracold bosonic atoms trapped in a double well created by an optical potential.  
The second can be implemented using quantum optical setups such as coupled resonant cavities. 
In this work we treat the former case, as atomic ensembles in optical lattices allow an high level of control on the parameters 
of the system. For instance the interaction between particles can be tweaked by using magnetic fields, thanks to Feshbach resonances 
and it is even possible to artificially simulate noise effects like dephasing and particle loss channels.
As a general rule, in all the above mentioned situations the resulting quantum states contain large amounts of entanglement that increase 
with the number of particles. This statement is not as trivial as it sounds, as there are situations where the entanglement 
does not increase with the particle number $N$, like for instance in the so-called NOON states, that contain the same 
quantity of entanglement as a Bell state, independent of $N$. Evidently, it is of great utility to perform a systematic study of the 
experimental strategies, among the multitude of possibilities realizable in modern laboratories, that allow to generate the 
largest amount of quantum resources. The detection and quantification of entanglement in these highly dimensional Hilbert spaces 
is still an unsolved problem for general mixed states, albeit the formidable research efforts spent in this direction during the last two decades. 
The difficulty is mainly caused by the presence of bound entanglement that eludes the typical separability tests, like the Peres' criterion.   
Fortunately, the two-wells problem where bosonic atoms occupy the first band is defined in a subspace of the whole 
Hilbert space, where quantum states are written in the so-called pair basis, that do not present such difficulties. 
Thanks to very recent theoretical achievements\cite{RoMoGe2013}, the entanglement in pair basis states can be safely detected 
and measured through the negativity. Moreover, entanglement of formation (EoF), i.e. the minimal amount of quantum physical resources 
needed to build the state, is estimable through some tight lower bounds. 

This article is structured as  follows. A first section introduces the system, illustrates the model that we use to describe it and 
explains the entanglement measures used. 
In the following, we consider the situation where the system is embedded in a thermal bath and we explore the effects of the variation of the 
jumping probability with respect to the inter-particle interaction strength at different temperature and for 
different particle number trapped in the potential. 
Then, we induce a quenched dynamics where a the hopping amplitude between the two wells is varied, showing the strategies 
that can be exploited to enhance the entanglement of the system. 
Later we study how detrimental effects like dephasing and single particle loss affect the coherence of the system 
in two circumstances: i) when it is initially prepared in its fundamental state and ii) in the presence of quenching dynamics. 
Finally, we calculate a lower bound for the EoF of the system.

\begin{figure}[pb]
\centerline{\includegraphics[width=7cm]{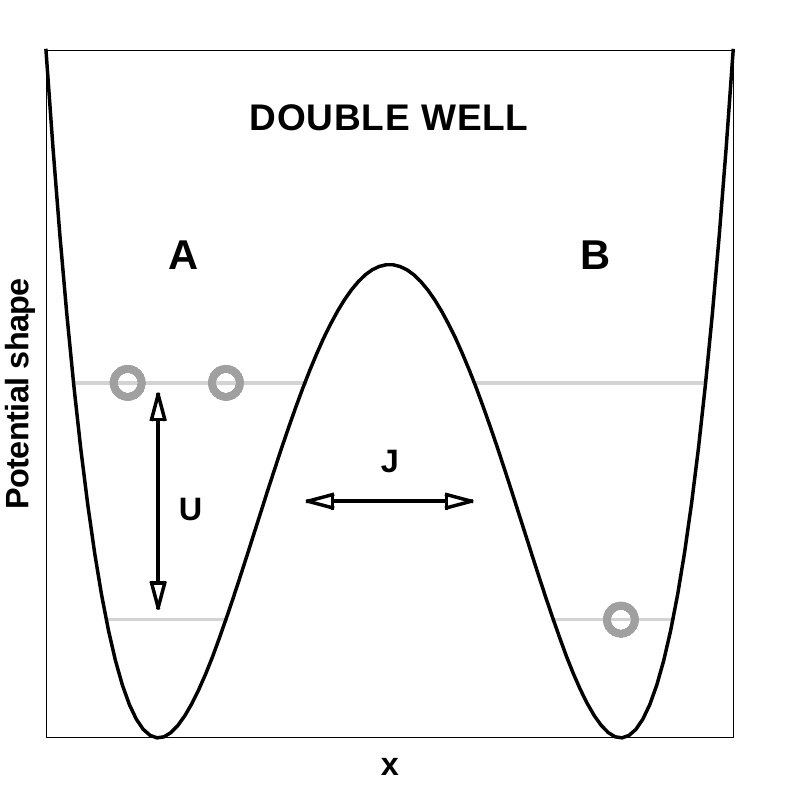}} 
\vspace*{8pt}
\caption{\label{fig:DW}Spatial profile of the double well potential. Bosons can move between A and B with hopping amplitude $J$ and interact via 
two-body collisions proportional to $U$.}
\end{figure}

\section{The model}
Consider a definite number $N$ of bosonic particles that are trapped in a double well potential. 
Assuming that the energies involved do not exceed the gap between first and second energetic bands, we can safely assume that 
only the first band is occupied. 
Then, it is possible to model the double well system as a really simple lattice formed by only two lattice sites $A$ and $B$, 
as sketched in figure~ \ref{fig:DW}. 
In this case, considering also the interactions, it is possible to use a simplified (restricted to two sites) version of the Bose-Hubbard Model.\cite{BlDaZw2008} 
The Bose-Hubbard Hamiltonian $\hat{H}$ is made up of two parts,
\begin{equation}
\hat{H}\left(J,U\right)=-J\hat{K}+U\hat{O}	\label{Bose-Hubbard}
\end{equation}
where
$\hat{K}=\hat{b}_{A}^{\dagger}\hat{b}_{B}+\hat{b}_{A}\hat{b}_{B}^{\dagger}$ describes the hopping of bosons between the two sites, as 
$\hat{b}_{\alpha}^{(\dagger)}$ is the annihilation (creation) operator of a particle in the site $\alpha \in \left\{ A,B\right\}$. 
The second term describes the the on-site interaction through the operator 
$\hat{O}=\sum_{\alpha\in\left\{ A,B\right\} }\hat{n}_{\alpha}(\hat{n}_{\alpha}-1)/2$, with 
$\hat{n}_{\alpha}=\hat{b}_{\alpha}^{\dagger} \hat{b}_{\alpha}$ the number operator. 
The physics is governed by two experimentally tunable parameters: the hopping amplitude $J$, related to the optical lattice intensity,  
and $U$, proportional to the two-body scattering length\cite{BlDaZw2008}. 
This latter quantity is tweakable through the mechanism of Feshbach resonance, which is one of the main peculiarity of cold atomic experiments. 
Owing to the coherent hopping between the two sites, a large amount of entanglement is espected to be generated. Its detection and quantification require the use of suitable measures.    

\subsection{Entanglement measures}
We choose the natural basis where the state corresponding to having $n$ particles in the well $A$ is $\left|n\right\rangle_{A} $. 
Since the total number of particles is $N$, then the $B$-restricted state will be $\left|N-n\right\rangle_{B} $. 
Therefore, the basis for the global state is given by the tensor product $\left|n\right \rangle_{A} \otimes \left|N-n\right\rangle_{B}$  
that we denote in short as $\left|n, N-n\right \rangle$, from now on. 
The structure of these states belongs to the so-called pair basis states, i.e. states of a bipartite system 
in which each of the basis states of one of the two subsystems is univocally associated to only one basis state of the other subsystem. 
Once the density matrix $\rho$ of the system is known, a typical indicator used to measure the entanglement, 
despite not being a good measure in general, is the negativity 
$\mathcal{N}\left(\rho\right)= \left(\left \Vert \rho ^{T_{A}} \right \Vert -1\right)/2$ 
where $\rho^{T_{A}}$ stands for the partial transpose with respect
to subsystem $A$ and $\left\Vert G\right\Vert _{1}=\mathrm{Tr}[\sqrt{GG^{\dagger}}]$ is the trace norm.
However, is has recently been shown that for pair basis states the negativity becomes a suitable entanglement measure and has the property of being discriminant\cite{RoMoGe2013}, since it is zero if and only if the state is separable. 
The negativity is easily computable, and for a pair basis state $\rho$ it becomes simply 
\begin{equation}
\mathcal{N} \left(\rho \right)=\sum_{i<j} \left\vert \rho_{i,j}\right \vert ,  \label{PB-neg}
\end{equation}
namely the sum of the absolute values of the off-diagonal elements. 

In this article we consider various situations of the many-body bosonic problem in a double well potential and characterize its entanglement. 
Taking advantage of the linear scaling of the Hilbert space dimension with $N$ (at variance with many other 
many-body problems that scale exponentially), we can efficiently diagonalize $\hat{H}$ and calculate the whole spectrum, 
even when the number of particles is relatively high. At the same time, also the eigenvectors are obtainable with arbitrary precision. 
Thanks to the above properties, the numerical calculation is done by using an ordinary computer, with modest resources of memory and time.  
The exact solution of the problem not only allows to get the ground state, but also to study its thermodynamics, 
and its time evolution after a change of external parameters both for the isolated system and in presence of dissipative processes.

\section{Thermal Bath}
Considering the whole system in equilibrium with a thermal bath, the density operator can be evaluated 
using the canonical ensemble\cite{PaBe1972,Ke1987} 
\[
\rho=\frac{e^{-\beta\hat{H}}}{\textrm{Tr}[e^{-\beta\hat{H}}]}
\]
which is manifestly a trace-one Hermitian matrix.
Since we know the spectrum of $\hat{H}$ we are able to evaluate the density matrix for such systems, given the temperature $T$ and the 
parameters $J$ and $U$. After this, it is easy to compute its entanglement level through the negativity $\mathcal{N}(\rho)$. 
In particular, it is interesting to see how the negativity changes by varying the two parameters $J$ and $U$.
Figuring out possible experiments, we choose to fix the interaction to $U=1$ and varying the hopping $J$, reducing 
to the sigle parameter $J/U$. 
Clearly, when the kinetic parameter is set to zero there is no interaction between the two wells, therefore no entanglement is generated. 
On the other hand, when $J/U$ tends to infinity, the kinetic part dominates. 
When we want to estimate the limit entanglement at low temperatures and 
$J\gg U$, we can consider the negativity of the ground state at $U=0$.
Using the creation and annihilation operators in second quantization it is possible to build the ground state for $U=0$, 
which constitutes the Bose-Einstein condensate (BEC) state spread in the two wells. This state can be written, for a system of $N$ 
particles, as follows:
\begin{equation}
\left|\Phi^{BEC}\right\rangle =\frac{1}{\sqrt{2^{N}N!}}\left(\hat{b}_{A}^{\dagger}+\hat{b}_{B}^{\dagger}\right)^{N}\left|0,\,0\right\rangle =\frac{1}
{2^{\frac{N}{2}}}\sum_{k=0}^{N}\frac{\sqrt{N!}}{\sqrt{k!}\sqrt{(N-k)!}}\left|N-k,\, k\right\rangle \label{BEC-state}
\end{equation}
with negativity  
\begin{equation}
\mathcal{N} \left(\rho_{BEC} \right)=\frac{1}{2^{N}} \sum_{k'<k=0}^{N} \sqrt{{N \choose k}{N \choose k'}} .\label{BEC-neg}
\end{equation}
where $\rho_{BEC}=|\Phi^{BEC}\rangle\langle\Phi^{BEC}|$.

\begin{figure}[h]
\centerline{\includegraphics[width=\columnwidth]{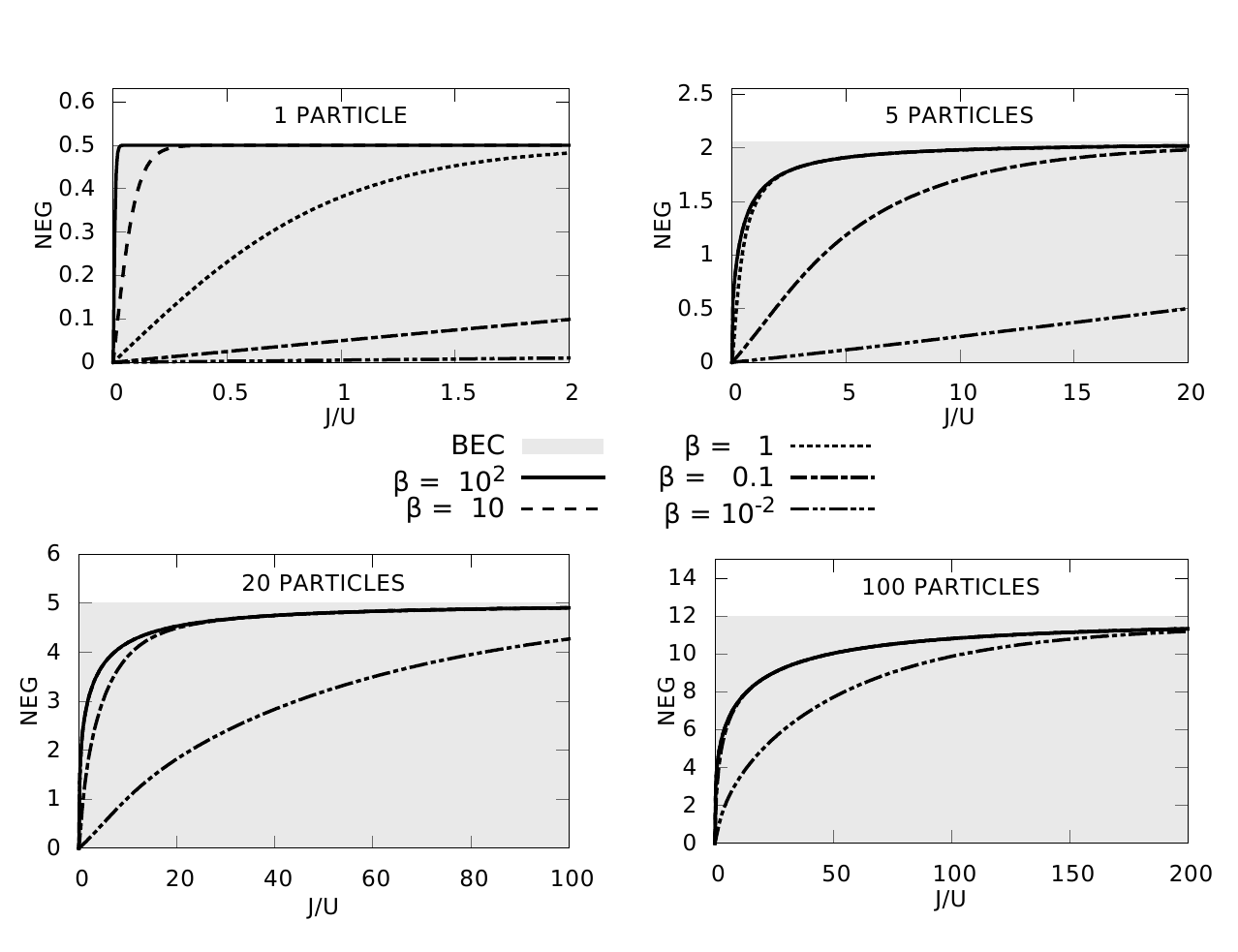}} 
\vspace*{8pt}
\caption{\label{fig:TherBath}The negativity of $N$ bosons in a double well potential at finite temperatures $T$ as a function of the ratio $J/U$. As customary, we indicate $\beta=\left(k_{B}T\right)^{-1}$. The shaded region represent the interval between zero and the asymptotic limit evaluated in eq.(\ref{BEC-neg}) of the BEC state negativity.}
\end{figure}

The numerical analysis shown in figure~\ref{fig:TherBath} has been performed at various temperatures $T$ and number of particles $N$. All the 
curves in function of $J/U$ start from zero since at $J=0$ the tunneling phenomena are suppressed and therefore the correlations between the 
two wells vanish. For a given $N$, the upper bound of the negativity is asymptotically attained at $J\gg U$ with the state $\Phi^{BEC}$. The detrimental effect of temperature over entanglement slows down the approach to the asymptotic limit $J/U\to\infty$.  
Figure~\ref{fig:TherBath} shows that increasing $N$ mainly induces two effects: i) the asymptotic limit grows, and ii) the upper 
bound is achieved at slower rate, even for very small temperatures. 
For $N=1$ the negativity approaches the asymptotic limit already at small $J/U$ for $\beta\ge5$, while for a 5-particle system this happens 
at $J/U\simeq20$ and for higher particle number at greater values. Nonetheless, the entanglement level for $N=5$ gains  
greater entanglement for the same values of $J/U$ and $\beta\ge1$.
\begin{figure}[h]
\centerline{\includegraphics[width=0.63\columnwidth]{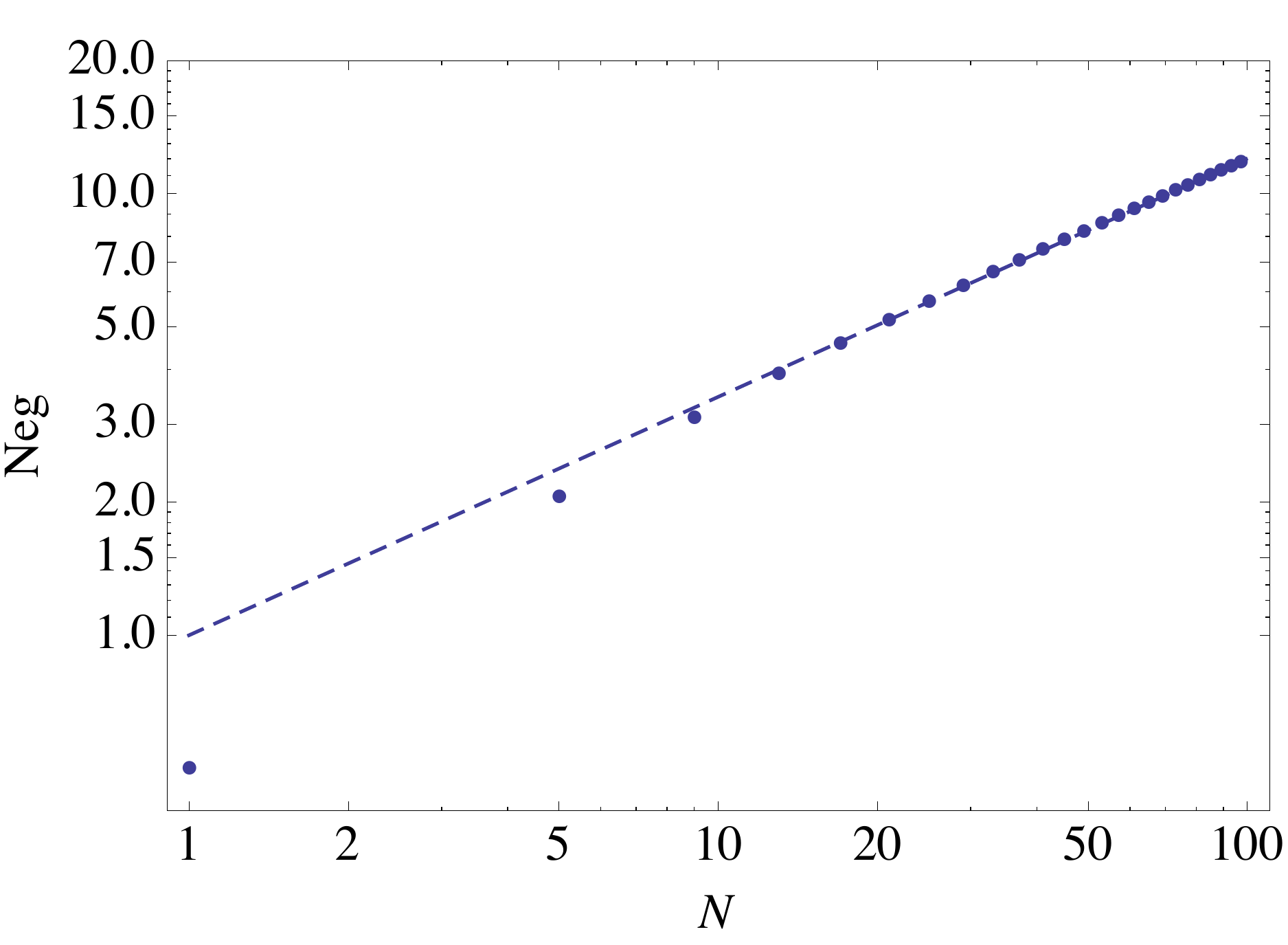}} 
\vspace*{8pt}
\caption{\label{fig:BEC-neg} 
The negativity eq.(\ref{BEC-neg}) of the BEC state in eq.(\ref{BEC-state}), namely the ground states obtained at $U=0$ and $J\neq0$. The curve as a function of the particle number $N$, is well fitted by a power law, as evidenced by the log-log plot.}
\end{figure}

In order to appreciate the growth of BEC-state entanglement by increasing the particle number, 
we plot $\mathcal{N} \left(\rho_{BEC} \right)$ as a function of $N$, as shown in figure~\ref{fig:BEC-neg}. 
Interestingly, the negativity grows as $\mathcal{N}(N)\approx N^{\alpha}$ with an exponent estimation $\alpha\sim 0.54$.


\section{Coherent Quenching Dynamics}
In this section we study the unitary temporal evolution after a quenching process, namely a sudden change of some parameters in the system Hamiltonian. In this specific case, we simply consider an instantaneous modification of the potential profile, causing a change in the hopping parameter $J$. This is experimentally feasible due to the high control obtainable in optical lattice. 
We assume that the system is initally prepared in the ground state of the Hamiltonian $\hat{H}_{0}(J_{0},U_{0})$. 
At time $t=0$ such state is evolved according to the Hamiltonian $\hat{H}_{e}(J_{e},U_{e})$. 
We consider the situation where the interaction parameter is left unchanged $U_{0}=U_{e}=1$, while the kinetic 
parameter is changed from the value $J_{0}=0.1$ to the value $J_{e}=1$. The increase of hopping rate is simply obtained by modifying 
the potential profile with a smaller barrier.
The initial ground state $\left|\psi_{0}^{0}\right\rangle$ can be expressed as a superposition of the eigenstates of the evolution 
Hamiltonian $\left\{ \left|\psi_{i}^{e}\right\rangle\right\} _{i=0}^{N}$
\begin{equation}
 \left|\psi_{0}^{0}\right\rangle =\sum_{i=0}^{N}c_{i}\left|\psi_{i}^{e}\right\rangle. \label{psi_o}
\end{equation}
Its evolution can be performed by using $\hat{H}_{e}$ as the generator of time translations: 
\begin{equation}
e^{-i\hat{H}_{e}t}\left|\psi_{0}^{0}\right\rangle =\sum_{i=0}^{N}e^{-iE_{i}^{e}t}c_{i}\left|\psi_{i}^{e}\right\rangle \equiv\sum_{i=0}^{N}c_{i}\left(t\right)
\left|\psi_{i}^{e}\right\rangle, \label{psi_t}
\end{equation}
where $\hat{H}_{e}\left|\psi_{i}^{e}\right\rangle =E_{i}^{e}\left|\psi_{i}^{e}\right\rangle$ and we have chosen units wth $\hbar=1$. 
The initial state of the system is no more an eigenstate of 
the Hamiltonian, hence the coefficients $c_{i}(t)$ start oscillating at various frequencies depending on the eigenvalues of the Hamiltonian $
\hat{H}_{e}$. Again, the complete numerical knowledge of the spectrum allow an exact calculation of the density matrix under evolution. 

After the quench we expect that the negativity of the density matrix starts changing with time. We observe this variation by evaluating the negativity $\mathcal{N}(\rho(t))$ at each time $t$.
In order to set the time-scale, we observe that after having set to $U$ the energy units in the Bose-Hubbard model Hamiltonian (\ref{Bose-Hubbard}), the parameter $\tau=1/U$ (remember that $\hbar=1$) sets the time unit.

Forcing the system through a stronger kinetic parameter causes an immediate gain of energy, 
\begin{eqnarray}
\langle \Delta E\rangle  _{\psi_{0}} 
        = \left\langle \psi_{0}\right|\hat{H}_{e}-\hat{H}_{0}\left|\psi_{0}\right\rangle \nonumber 
	= \left\langle \psi_{0}\right|\left[-(J_{e}-J_{0})\hat{K}\right]\left|\psi_{0}\right\rangle =-\Delta J\langle \hat{K}\rangle _{\psi_{0}} 
	\label{energy}
\end{eqnarray}
since for $J_{0}<J_{e}$ the kinetic operator $\hat{K}$ is positive. At later times, the evolution operator commutes with the new 
Hamiltonian $\hat{H}_{e}$ and the energy turns to be constant of the motion again. Hence, the energy of the system decreases from its initial value at time $t_{0}=0$.

\begin{figure}[h]
\centerline{\includegraphics[width=\columnwidth]{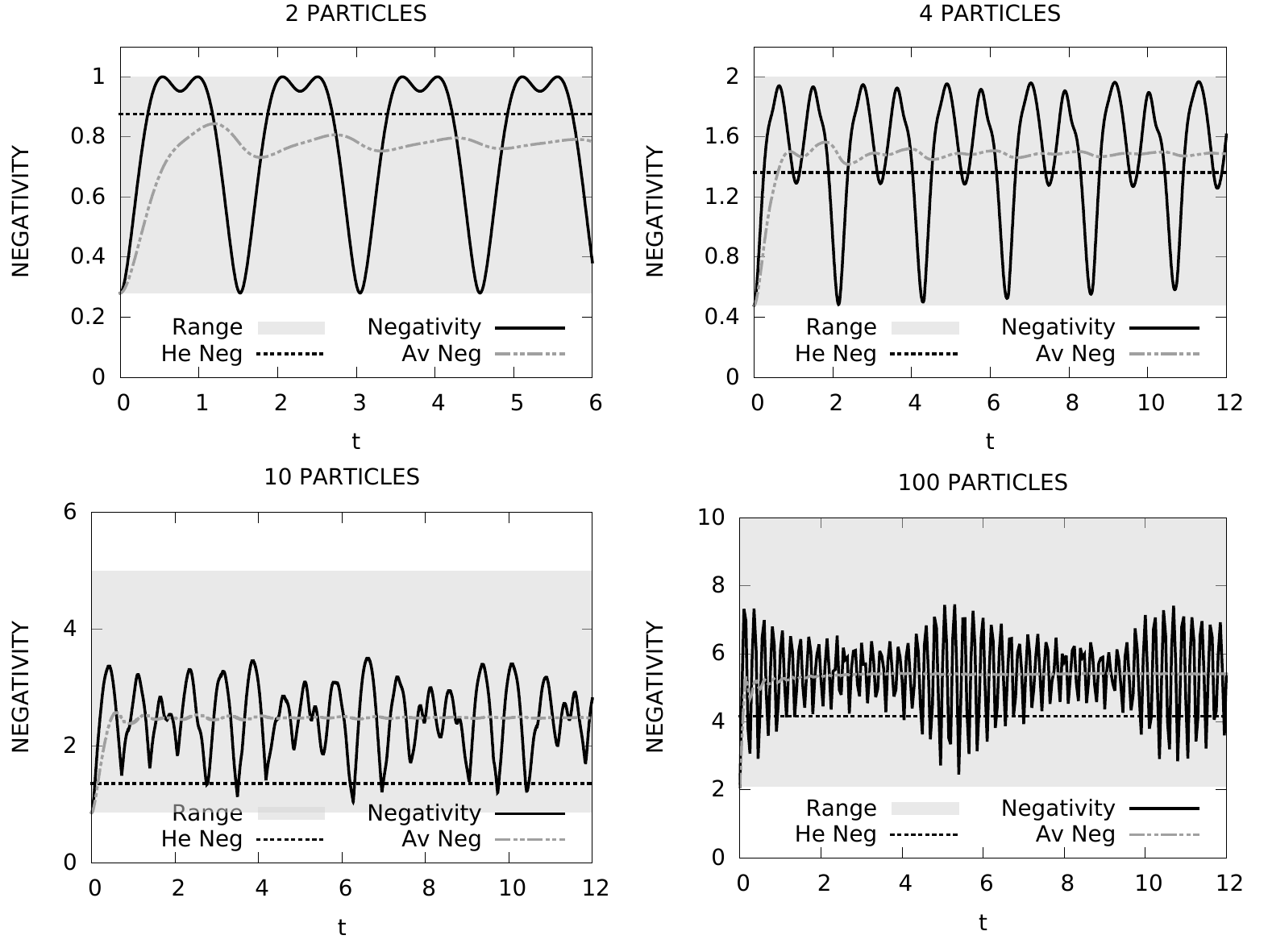}} 
\vspace*{8pt}
\caption{\label{fig:CohQuen}Coherent evolution of the negativity $\mathcal{N}$ for some number of particles in a double well potential after the  quench process. It is evident how the time-averaged mean $\bar{\mathcal{N}}$ (dot-dashed line) quickly stabilizes in 
time between the starting value and the maximum feasible negativity $\mathcal{N}_{\textrm{Max}}=\frac{N}{2}$. The shaded region labelled as 
``Range" represents the interval between the initial negativity outcome and $\mathcal{N}_{\textrm{Max}}$. The dashed line marks the negativity for the ground state of $\hat{H}_e$. The time scale is expressed in units of $1/U$ ($\hbar=1$).}
\end{figure}

At this stage, we do not need to perform a numerical solution of the Liouville equation $\dot{\rho}=-i[ \hat{H},\rho ]$, but we solve it in terms of the evolution operator $e^{-i\hat{H}_{e}t}$. Since $\hat{H}_{e}$ is in general a large matrix, the determination 
of the evolution operator can be efficiently made only by numerical diagonalization algorithms. 
The calculation of the negativity is presented in figure~\ref{fig:CohQuen}, for several values of the particle number. 

The numerical results tell that for any $N$ the entanglement is enhanced (except for a numerable union of zero-measure sets) 
with respect to its initial value and its time-integrated mean value stabilizes very quickly (see figure~\ref{fig:CohQuen}).
For every considered case it is possible to identify temporal intervals for which the negativity is higher than the value that it would have in 
the ground state of the final Hamiltonian (dashed horizontal lines in figure~\ref{fig:CohQuen}). That means that the sudden quench gives 
enhanced results with respect to an adiabatic transformation that interpolates between the two ground states. 
The particle number $N$ of the system plays a twofold role. On the one hand, as $N$ increases the entanglement grows, up to leading its time-integrated average $\bar{\mathcal{N}}(t)=t^{-1}\int_0^t dt' \mathcal{N}(t')$ above the ground state negativity of $H_e$. On the other hand, if for low $N$'s values the negativity approaches the maximum feasible value $\mathcal{N}_{\textrm{Max}}=N/2$, this effect no further appears for higher particle numbers. 

\section{Open Systems Dynamics}
Unlike closed systems, where the time evolution is governed by the Liouville equation, in open systems a unified treatment describing the temporal evolution for each subsystem does not exists.\cite{NiCh2000,BrPe2002} 
In this work, we assume that the system evolution is Markovian, hence to perform temporal evolution of the system we use the Lindblad equation\cite{Li1976}
\begin{equation}
\dot{\rho}=-i[H,\rho]+\sum_{k>0}(L_{k}\rho L_{k}^{\dagger}-\frac{1}{2}\{L_{k}^{\dagger}L_{k},\rho\}),\label{lindblad}
\end{equation} 
where $L_k \rho L_k^\dagger$ indicates one of the possible quantum jump operators. Specifically, we 
consider two different decoherent processes: i) dephasing and ii) particle loss. First, we will observe the effects of 
these decoherence on the entanglement of the ground states for the Bose-Hubbard Hamiltonian. 
Later, we consider the dynamics of the entanglement quenching of these two open systems.

\subsection{Phase Decoherence}
A dephasing channel represents the case in which the system is coupled in a particle-number conservative way to the environment. As an 
example, it can model the case in which the system undergoes to a sequence of destructive measurements. 
In condensed matter physics and in solid state physics, this type of decoherence is frequently caused by the interaction between electrons and phonons.\cite{Sc1976,Hu2011,TiZhFi2014} 
While in optical lattices the degradation due to phase decoherence is in general very low, it can be purposely induced for quantum simulation reasons; for instance, in order to study the effects of phononic interactions in a crystal.
In this case, the phononic disturbance is simulated introducing a condensate that does not feel the optical lattice, but interacts with the trapped particles. The interaction can then be tuned using tweaking the inter-species scattering length through Feshbach resonances. The range of the possible values for the rate of decoherence is very large. 

The associated quantum jump operators for the double well problem in the second quantization formalism are:
\begin{equation}
\Gamma_{k}=\gamma_{k}\hat{n}_{k},\; k=A,\, B
\end{equation} 
where $\gamma_{k}$ indicates the decoherence rate at site $k=\mathrm{A}, \mathrm{B}$. 
For the sake of simplicity, in the following we will assume $\gamma_{A}=\gamma_{B}=\gamma$. 
Hence, the Lindblad master equation takes the form:
\begin{eqnarray}
\dot{\rho}	=	\frac{1}{i}\left[\hat{H},\rho\right]
		+	\gamma\left[\hat{n}_{A}\rho\hat{n}_{A}+\hat{n}_{B}\rho
		\hat{n}_{B}-\frac{1}{2}\left(\hat{n}_{A}\hat{n}_{A}\rho+\rho\hat{n}_{A}\hat{n}_{A}+\hat{n}_{B}\hat{n}_{B}\rho+\rho\hat{n}_{B}\hat{n}_{B}\right)\right] \label{lindblad-dephasing}
\end{eqnarray}
Such equation, which can be cast in the form $\dot{\rho}=\mathcal{L}[\rho]$, is a first order differential equation that can be solved using standard numerical methods. 

\subsubsection{Ground state for the dephasing channel}
\begin{figure}[h]
\centerline{\includegraphics{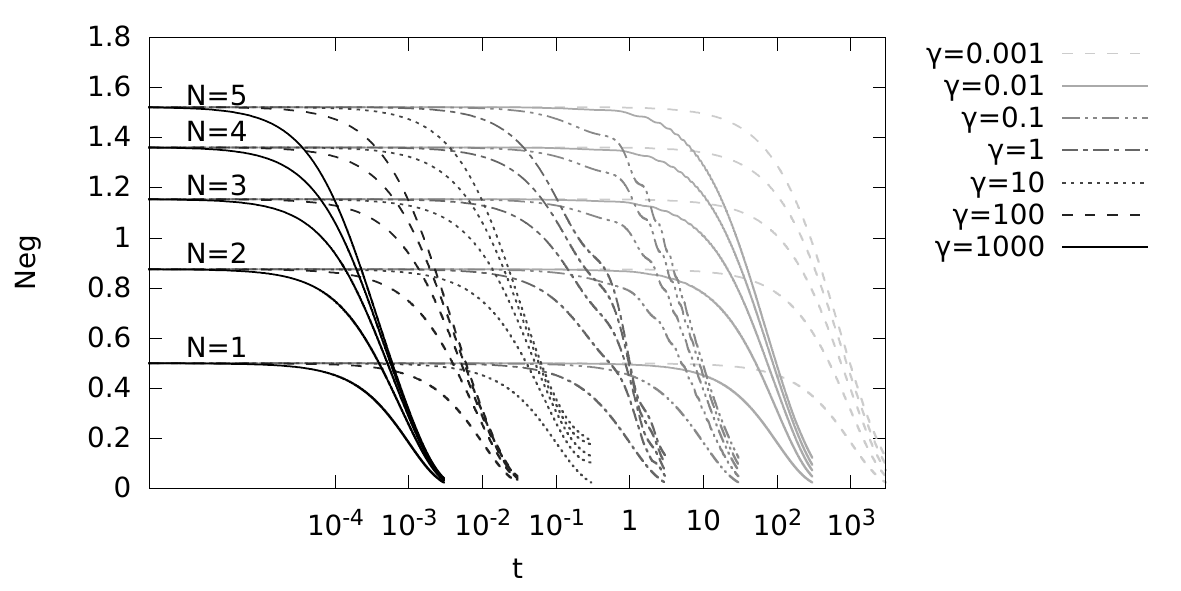}} 
\vspace*{8pt}
\caption{\label{fig:DephGS}The initial ground state of a ($N=1,2,3,4,5$)-particle Bose Hubbard double-well Hamiltonian with 
$J=1$ and $U=1$ is time-evolved in a dephasing channel at different values of  $\gamma$.}
\end{figure}

First we consider the following case: the system is initially in the ground state of a Bose-Hubbard double well Hamiltonian of given ratio $J/U=1$ 
(respectively between the kinetic and on-site interaction parameters). Successively the system is evolved in a decoherent channel with rate $\gamma$ (for numerical values, see figure~\ref{fig:DephGS}). We expect (for $t\gg \gamma^{-1}$) the matrix to 
asymptotically collapse into a diagonal matrix. 
Since the amount of entanglement is well described by its negativity and this quantity takes the simple 
form $\mathcal{N}(\rho)=\sum_{i<j}\left|\rho_{ij}\right|$ in this case (i.e. the sum of the modulus of the off-diagonal matrix elements), 
the negativity is expected to approach zero asymptotically for long times. 
The range of possible decoherence ratios is very large, it can be intentionally tweaked in quantum simulation as stressed in the previous 
section. 

\subsubsection{Quenching in a phase dumping channel}
The system is initially prepared in the ground state of a double-well 
Bose-Hubbard Hamiltonian of kinetic and on-site interaction parameters respectively equal to $J_{i}/U_{i}=0.1$. At time $t=0$, the system is left 
to evolve in a double-well with $U_{e}=U_{i}=J_{e}$. The difference from the coherent case is that now the system is subjected to phase 
decoherence with rate $\gamma$.

\begin{figure}[h]
\centerline{\includegraphics[width=\columnwidth]{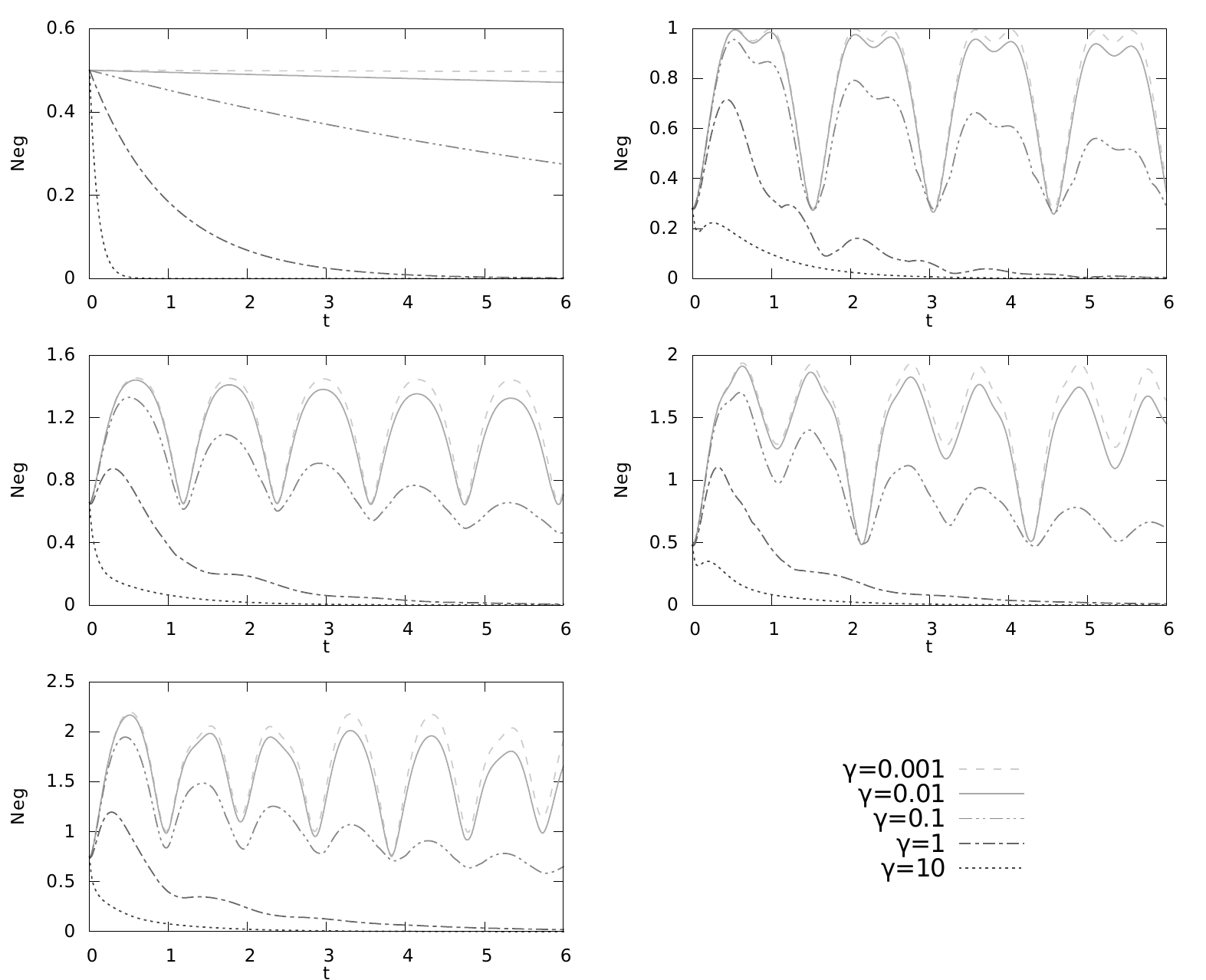}} 
\vspace*{8pt}
\caption{\label{fig:DephQ}Dephasing evolution in a quench dynamics. The ground state of a ($N=1,2,3,4,5$)-particle Bose Hubbard double-
well Hamiltonian with $J=1$ and $U=1$ is time-evolved in a phase dumping channel at different values of  $\gamma$. From one particle in the top left box increasing left-right top-bottom up to the 5-particle case in the bottom left box.}
\end{figure}

The negativity of the system is expected to oscillate even at high values, in accordance to the observations done in the coherent 
case for times $t\ll \gamma^{-1}$ except in case of strong decoherece regimes ($\gamma\ge 10$ as it can be seen in figure~\ref{fig:DephQ}). Then, even if still oscillating, $\mathcal{N}$ drops asymptotically to zero for times $t\gg\gamma^{-1}$.
Calculations show that the system does not oscillate if it contains only one particle, because the matrix elements of the 
eigenstates of such system possess equal modulus, hence equal negativity.
From numerical results, it can be observed that the entanglement remains 
above the threshold of its initial value for times $t\ll\gamma^{-1}$. 
The time at which the system crosses this threshold depends on $N$.

\subsection{Particle Loss}
In this section we assume that the system may lose particles with a certain probability for unit time. 
In physics of ultracold gases, dissipation effects naturally occur for bosonic atoms in the form of three-body losses.\cite{FeReSh1996,Pe2004} This can be interesting to study in some situations and, in the area of quantum simulation, one can even be interested in increase this dispersive process. For instance in Ref.~\refcite{RoRiCi2010} it is proposed a technique to enhance particle loss decoherence in a bosonic gas increasing the density of the gas.

A two-particle loss mechanism can be induced on an optical lattice using photo-association, irradiating the gas with laser pulses.
\cite{DuGaSyBaLeCiRe2009} 
This process cause the formation of biatomic molecules that, being no more sensible to the optical lattice, flow away from the trap.
Differently, here we study the effects of one-body losses, which are especially important in quantum-optical implementations, like optical cavities. 
In this case the decoherence rate is inversely proportional to the quality factor\cite{Ca2001,Ca2008} (also referred to as $Q$-value) 
$Q=\omega_{0}T_{\textrm{rt}}/I$, where $\omega_{0}$ is the resonant frequency of the cavity, $T_{\textrm{rt}}$ is the round trip time and $I$ is the fractional power loss. 
An equivalent expression for the quality factor is $Q=\omega_{0}/\Delta\omega$ where $\Delta\omega=\omega_{d}-
\omega_{0}$ and $\omega_{d}$ is the pumping laser frequency. The $Q$-value can vary in a wide range, 
depending on the experimental setup. 

In this latter situation, the quantum jumps are represented by the operators $\Gamma_{k}=\gamma_{k}\hat{b}_{k},\, k=A,B$ since the process of losing one particle in the well $k$ is determined by applying the annihilation operator $\hat{b}_{k}$. The coefficient $\gamma_{k}$ represents the 
decoherence rate, that is proportional to the probability of losing a particle in the well $k$. Even in this case, it will be considered the simple case 
$\gamma_{A}=\gamma_{B}=\gamma$. Hence the Lindblad equation takes the form
\begin{eqnarray}
\dot{\rho}=\frac{1}{i}\left[\hat{H},\rho\right]
+\gamma\left[\hat{b}_{A}\rho\hat{b}_{A}^{\dagger}+\hat{b}_{B}\rho\hat{b}_{B}^{\dagger}-\frac{1}{2}\left(\hat{b}_{A}^{\dagger}\hat{b}_{A}\rho+\rho\hat{b}_{A}^{\dagger}\hat{b}_{A}+\hat{b}_{B}^{\dagger}\hat{b}_{B}\rho+\rho\hat{b}_{B}^{\dagger}\hat{b}_{B}\right)\right]
\end{eqnarray}
Obviously, as for long times ($t\gg \gamma^{-1}$) the density matrix collapses into the element corresponding to empty system, the 
negativity vanishes asymptotically.

\subsubsection{Ground state in a Particle-loss channel}
Consider the case in which the eigenstate of a $N$-particle Bose-Hubbard Hamiltonian with $U=J=1$ is left to evolve into a particle-loss 
channel with a certain decoherence ratio $\gamma$. 
As the time goes on, the density matrix (initially populated only into the submatrix of order $N+1$ of the initial state) 
is expected to deplete in the regions of higher population in favor of less populated regions.

If the initial density matrix is sorted such that the basis states of the well $A$ appear in decreasing order, the density matrix will be structured with 
a set of square sub-matrices sorted in decreasing order from $N+1$ to $0$ situated along the diagonal. The first top-left matrix is the biggest and 
corresponds to the full system, while the last element at the bottom-right of the matrix corresponds to the empty system. 
All the other sectors of the matrix correspond to the quantum correlations present between two $M,\, P\le N$ populated systems. Obviously if the 
correlations between differently populated systems are strictly classical these regions must be identically zero. 
\begin{figure}[h]
\centerline{\includegraphics{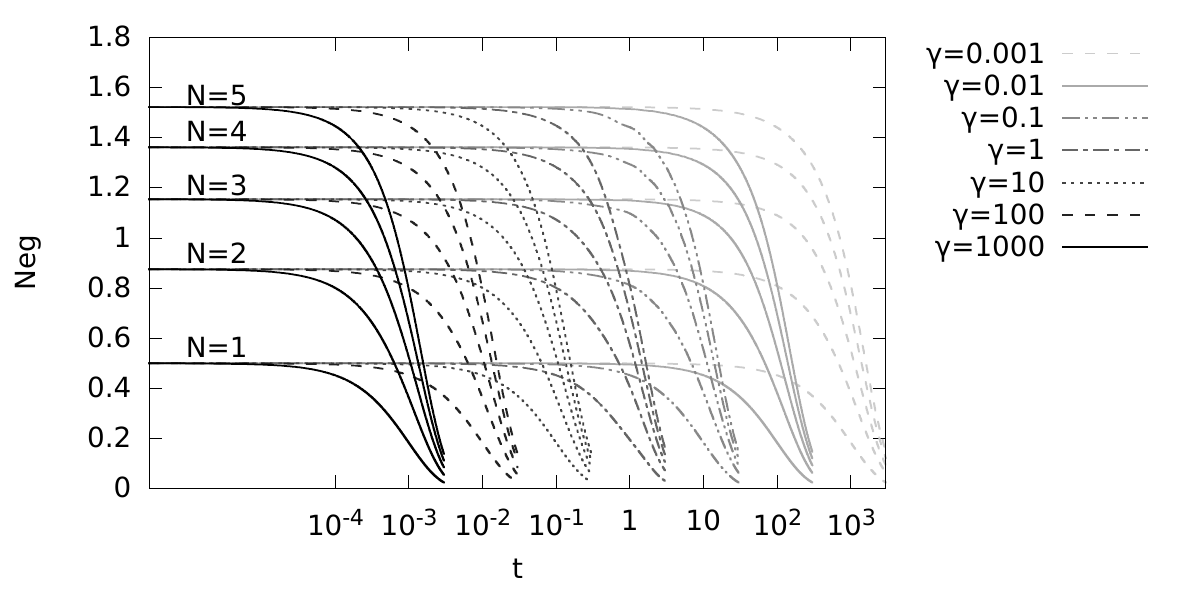}} 
\vspace*{8pt}
\caption{\label{fig:PLGS}Particle-loss evolution. The ground state of a ($N=1,2,3,4,5$)-particle Bose Hubbard double-well Hamiltonian with $J=1$ 
and $U=1$ is time-evolved in a dephasing channel at different values of  $\gamma$.}
\end{figure}
If this applies, then the density matrix will be a direct product of the $N+1$ density matrices that do no possess coherence from 
one to each other. Therefore, the negativity remains a suitable entanglement measure for this system and its value is equal to the sum of the 
negativities of the block density matrices. All these properties can be observed in figure~\ref{fig:PLGS}.

\subsubsection{Quenching in a particle-loss channel}
\begin{figure}[h]
\centerline{\includegraphics[width=\columnwidth]{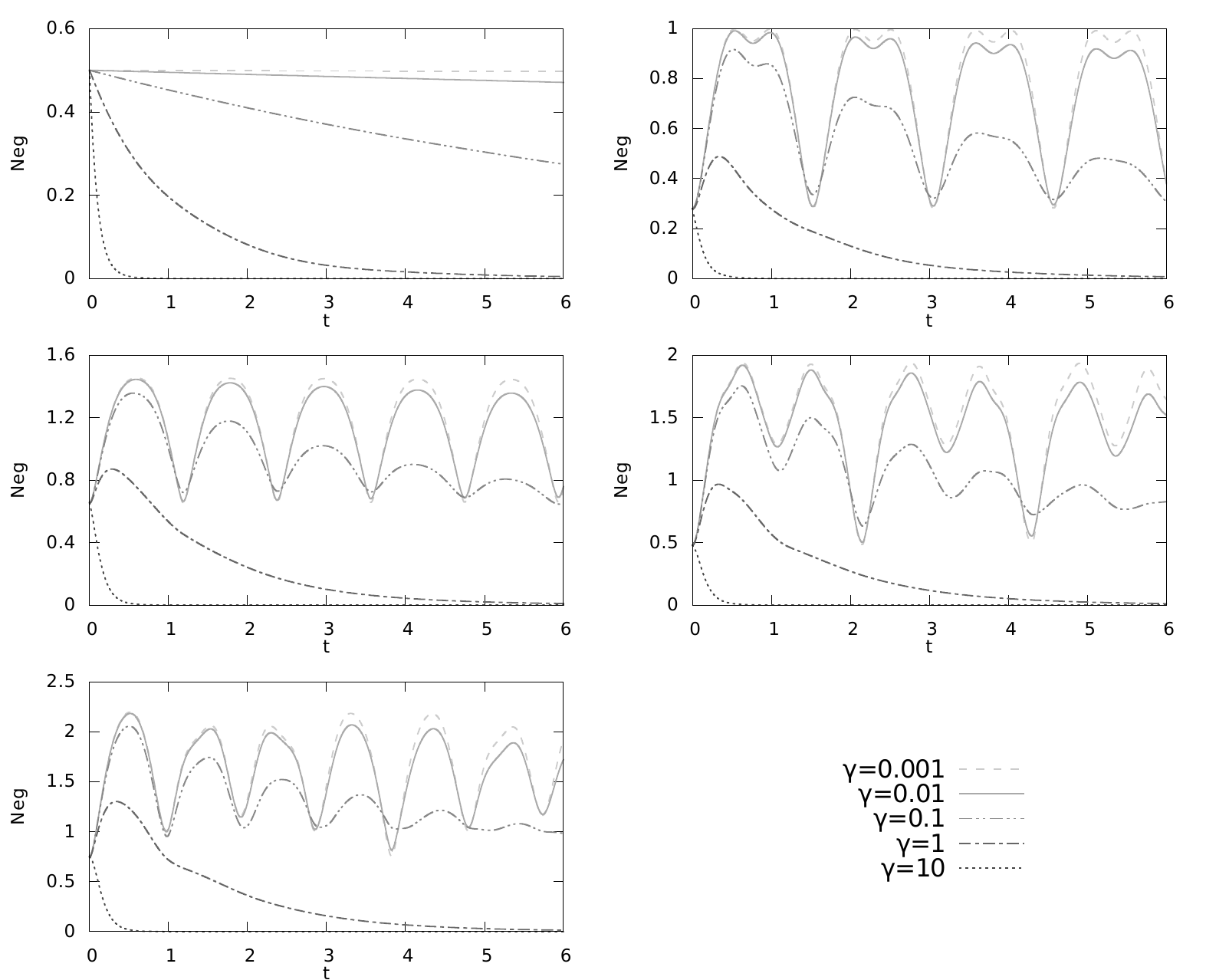}} 
\vspace*{8pt}
\caption{\label{fig:PLQ}Quenched particle-loss evolution. The ground state of a ($N=1,2,3,4,5$)-particle Bose Hubbard double-well Hamiltonian 
with $J=1$ and $U=1$ is time-evolved in a particle-loss channel at different values of  $\gamma$ in quenching dynamics. From one particle in 
the top left box increasing left-right top-bottom up to the 5-particle case in the bottom left box.}
\end{figure}

Analogously to the dephasing situation, here we consider the case where a quenching dynamics is evolved in presence of dissipation. Now, the dissipative process that affects the system is the particle-loss channel. We consider a system initially in the ground state of the Bose-Hubbard 
Hamiltonian of parameters ratio $J_{0}/U_{0}=1$ evolved in a decoherent double-well potential of parameters ratio $J_{e}/U_{e}=0.1$ with 
decoherence rate $\gamma$ (see figure~\ref{fig:PLQ}).
Since the wide range of possible values of $\gamma$ different regimes can be explored in this study.

\section{Lower bound of the entanglement of formation}

Since our system is described through pair basis states, it is possible to estimate lower bounds for the EoF 
as explained in Ref.~\refcite{RoMoGe2013}. 
This can be done by evaluating the maximum of three quantities $F$, $G$, and $s$ (whose definition will be given soon after), which are functions of the entries of the density matrix $\rho$. 

Before proceeding in the evaluation it is better to reorganize rows and columns of the density matrix, 
in decreasing order with respect to the quantity $\Gamma_{i}^{2}=\sum_{j\neq i}\left|\rho_{ij}\right|^{2}$, i.e. the sum of the 
square moduli of the off-diagonal elements of a given row $i$. 
Once the density matrix is reordered with $\Gamma_{1}\ge\Gamma_{2}\ge\cdots\ge\Gamma_{d}$
(indicating with $d$ the dimension of the system) $F(\rho)$ takes
the form
\begin{equation}
F(\rho)=-\sum_{i=1}^{d}\alpha_{i}^{2}(\mathbf{x})\log\alpha_{i}^{2}(\mathbf{x})\label{eq:F}
\end{equation}
where 
\[
\alpha_{1}^{2}(\mathbf{x})=\frac{1+\sqrt{1-4|\mathbf{x}|^{2}}}{2};\qquad\alpha_{i\neq1}^{2}(\mathbf{x})=\frac{|\mathbf{x}_{i}|^{2}}{\alpha_{1}^{2}},
\quad(i\le d)
\]
and $\mathbf{x}=\left(\rho_{12},\rho_{13},\cdots,\rho_{1d}\right)$
is the vector of the off-diagonal elements of the first row of $\rho$.
The quantity $F(\rho)$ has been rigorously proved to be a lower bound for the EoF of general mixed states 
written in a pair basis\cite{RoMoGe2013}. The second function, called $G(\rho)$ is given, analogously to $F(\rho)$,
as
\begin{equation}
G(\rho)=-\sum\alpha_{i}^{2}(\mathbf{y}_{i})\log\alpha_{i}^{2}(\mathbf{y}_{i})\label{eq:G}
\end{equation}
but 
\[
\alpha_{i}^{2}(\mathbf{y}_{i})=\frac{1}{2}\left(1-(-1)^{\delta_{i,1}}\sqrt{1-4|\mathbf{y}_{i}|^{2}}\right)
\]
and $\mathbf{y}_{i}=\{\rho_{ij},\, i\neq j\}$, namely we take all the off-diagonal elements of $\rho$, at variance with the case of $F$.   

Finally, the function $s(\rho)$ is already known in literature\cite{TeVo2000,ChAlSh2005} to be a lower bound of the EoF for arbitrary states.  
The function $s$ depends on $\rho$ only through a single parameter, that in the restricted family of pair states 
turns out to be the negativity $\mathcal{N}$. Explicitly, such a function $s(\mathcal{N})$ is found to be
\begin{equation}
s(\mathcal{N})=
\cases{
H_{2}(\gamma)+(1-\gamma)\log N, & $\mathcal{N}\in\left[0,\frac{3}{2}-\frac{2}{N+1}\right]$\cr
\frac{2\mathcal{N}-N}{N-1}\log N+\log(N+1), & $\mathcal{N}\in\left[\frac{3}{2}-\frac{2}{N+1},\frac{N}{2}\right]$
}\label{eq:Lower-bound}
\end{equation}
where 
$\gamma(\mathcal{N})=\frac{1}{d^{2}}[\sqrt{2\mathcal{N}+1}+\sqrt{(d-1)(d-2\mathcal{N}-1)}]^{2}$ and $H_{2}(\cdot)$ is the 
Shannon bipartite entropy. Summarizing, a lower bound for the EoF for pair basis states, is given by
\[
\max\{F(\rho),G(\rho),s(\rho)\} \label{eq:Max-FGs}
\]
using the expressions (\ref{eq:F}) (\ref{eq:G}) and (\ref{eq:Lower-bound}).

\subsection{Double well ground states EoF lower bound}
\begin{figure}[pt]
\centerline{\includegraphics[width=\columnwidth]{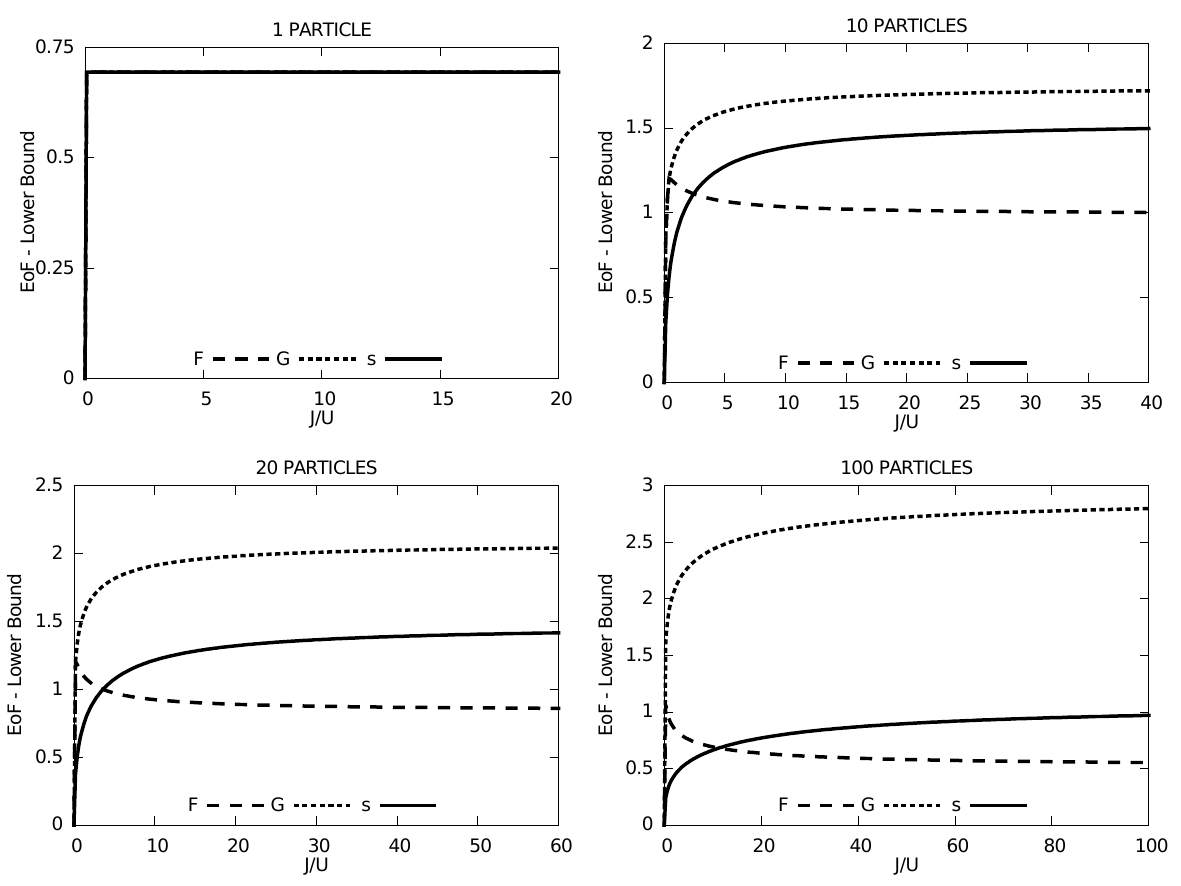}} 
\vspace*{8pt}
\caption{\label{fig:EoF}Lower bound for the EoF of the ground state of the Bose-Hubbard double well Hamiltonian. In the boxes is represented 
the behaviour of (\ref{eq:Max-FGs}) as a function of the ratio $J/U$ for 1, 5, 20 and 100 particles in order left--right--top--bottom.}
\end{figure}

In the following we analyze the EoF lower bound of the ground state of the double well Bose-Hubbard Hamiltonian, varying the ratio $J/U$. This 
study is realized for different numbers of bosons trapped in the double well potential and the graphical results can be seen in figure~
\ref{fig:EoF}.
Obviously, the lower bound for the EoF (\ref{eq:Max-FGs}) has outcome zero for $J/U=0$ in each case considered. 
In the 1-particle case, the EoF is exact and has a sudden jump as soon as $J/U\neq0$. 
For higher particle numbers, the considered lower bounds grow in a smoother way. 
Curiously, as it can be observed in figure~\ref{fig:EoF}, the behaviour of the $F$-function shows a sharp peak for intervals $\left(0,\delta\right]$ in the $J/U$ domain, where $\delta >0$. After the peak, the estimated lower bound of EoF decreases as $J/U$ grows, while the entanglement is expected to increase. This drawback is due to the nearly uniform distribution of Schmidt coefficients for the states that are close to the BEC one ($J/U\to \infty$), a situation where $F$ is known to work less better that the other indicators. We report that in other cases, like the infinite dimensional twin-beam states emerging from down-converted photons in quantum optics, the distribution of Schmidt coefficients is not so spread and $F$ gives a much better estimation of the EoF\cite{RoMoGe2013}. Anyway the most striking feature that emerges from figure~\ref{fig:EoF} is that for a wide range of couplings the function $G$ yields the better lower bound of EoF as compared with the values of $F$ and $s$. 
Such discrepancy is more and more evident as the number of particles is increased. 

\section{Conclusions}

Using some recent results on the estimation of entanglement in pair basis states\cite{RoMoGe2013}, 
we have explored a system capable to generate a huge quantity of quantum information resources: 
bosonic particles in a double-well potential. 
The Hamiltonian is readily implementable with the modern experimental techniques both in cold atomic ensembles trapped 
in optical lattices and in quantum optical setups of coupled cavities. These systems could serve as prototypical devices for 
generating non classical resources for implementing protocols that require large amounts of entanglement. 
By using the numerical solution coming from exact diagonalization, we have studied the system in many situations, including 
the quenched dynamics after the sudden lowering of the barrier between the two wells. This latter strategy turns out to be one of 
the most promising for obtaining the largest negativities. 
We have also studied the stablity of the produced entanglement under the detrimental effect of temperature, decoherence and particle losses.  
This work suggests and stimulates the investigation of other engineered procedures optimized to generate entanglement 
starting from atomic condensates, i.e. objects known to be poor of quantum correlations.   

\section*{Acknowledgements}
We acknowledge the Compagnia di San Paolo for support.

\end{document}